\documentclass[a4paper,fleqn,usenatbib]{mnras}
\usepackage{bm}
\usepackage{newtxtext,newtxmath}
\usepackage[T1]{fontenc}
\usepackage{ae,aecompl}
\usepackage{graphicx}	
\usepackage{amsmath}	
\usepackage{amssymb}	
\usepackage{here}

\newcommand{\fref}[1]{Fig.~\ref{#1}}
\newcommand{\Msun}{\hbox{$M_{\sun}$}}
\newcommand{\Mhost}{\hbox{$M_{\rm host}$}}
\newcommand{\Mratio}{\hbox{$M_{\rm ratio}$}}
\newcommand{\Fnumber}{\hbox{$F_{\rm number}$}}
\newcommand{\Fmass}{\hbox{$F_{\rm mass}$}}

\renewcommand{\vec}[1]{\hbox{$\boldsymbol{\mathit{#1}}$}}
\newcommand{\efilament}{\vec e_{\rm filament}}
\newcommand{\emajor}{\vec e_{\rm major}}
\newcommand{\J}{\vec J}
\title[Impact of filamentary accretion on halo structures]{The impact of filamentary accretion of subhaloes on the shape and orientation of haloes}

\author[Yu Morinaga et al.]{
Yu Morinaga 
and Tomoaki Ishiyama$^{1}$\thanks{E-mail: ishiyama@chiba-u.jp}
\\
$^{1}$Department of Applied and Cognitive Informatics, Chiba University, 1-33, Yayoi-cho, Inage-ku, Chiba 263-8522, Japan\\
$^{2}$Institute of Management and \textcolor{black}{Information} Technologies, Chiba University, 1-33, Yayoi-cho, Inage-ku, Chiba 263-8522, Japan\\
}
\date{Accepted XXX. Received YYY; in original form ZZZ}
\pubyear{2018}

\begin{document}
\label{firstpage}
\pagerange{\pageref{firstpage}--\pageref{lastpage}}
\maketitle

\begin{abstract}
Dark matter haloes are formed through hierarchical mergers of smaller
haloes in large-scale cosmic environments, and thus anisotropic
subhalo accretion through cosmic filaments have some impacts on halo
structures.  Recent studies using cosmological simulations have shown
that the orientations of haloes correlate with the direction of
cosmic filaments, and these correlations significantly depend on the
halo mass.  Using high-resolution cosmological $N$-body
simulations, we quantified the strength of filamentary subhalo accretion
for galaxy- and group-sized host haloes ($M_{\rm host}=5\times10^{11-13}M_{\sun}$)
by regarding the entry points of subhaloes as filaments
and present statistical studies that how the shape and orientation of host haloes at
redshift zero correlate with the strength of filamentary subhalo accretion.
We confirm previous studies that found the host halo mass dependence of the alignment
between orientations of haloes and filaments.
We also show that, for the first time, the shape and orientation of
haloes weakly correlate with the strength of filamentary
subhalo accretion even if the \textcolor{black}{host} halo masses are the same.
Minor-to-major axis ratios of haloes tend to decrease as
their filamentary accretion gets stronger.  Haloes with highly
anisotropic accretion become more spherical or oblate, while haloes with
isotropic accretion become more prolate or triaxial.  For haloes
with strong filamentary accretion, their major axes are
preferentially aligned with the filaments, while their angular
momentum vectors tend to be slightly more misaligned.
\end{abstract}

\section{Introduction}
According to the standard cosmological model, dark matter haloes are
assembled via the hierarchical mergers of a number of smaller haloes
\citep{1978MNRAS.183..341W}.  Haloes reside in large-scale cosmic
environments so-called "cosmic web" \citep{1996Natur.380..603B}, which
are classified as voids, sheets, filaments and clusters.  We can
recognize these environments by wide-field galaxy surveys,
such as the Sloan Digital Sky Survey
\citep{2000AJ....120.1579Y}.  Some properties of galaxies such as
colour, age, size, and luminosity function of satellite galaxies depend
on cosmic environments \citep[e.g.,][]{2011MNRAS.413.2288M,
  2015ApJ...800..112G,2015MNRAS.450.2727T,2017MNRAS.466.1880C},
indicating that large-scale environments are responsible for formation histories of haloes
and galaxies embedded in them.

Cosmological simulations have been suggesting that
assembly histories of haloes depend on the environments around them
\citep[e.g.][]{2007MNRAS.375..489H, 2007MNRAS.381...41H,
  2007ApJ...654...53M}, and such
environmental effect would characterize some
properties of haloes such as the shape, orientation, angular momentum and spin
\citep[e.g.,][]{2006ApJ...652L..75P,2007MNRAS.375..489H, 2007MNRAS.381...41H, 2009ApJ...706..747Z,
2011MNRAS.413.1973W,2011MNRAS.416.1377V,2012MNRAS.421L.137L,
  2013MNRAS.428.2489L,2013ApJ...762...72T,2015ApJ...813....6K,24017MNRAS.466.3834L,2017ApJ...848...22X,
2017MNRAS.468L.123W,2018MNRAS.481..414G,2019JCAP...10..020O,2019arXiv190803467L}.
For example, major axes of host haloes tend to be preferentially
aligned with the directions of filaments \citep{2007MNRAS.381...41H,
  2009ApJ...706..747Z, 2013MNRAS.428.2489L, 2018MNRAS.481..414G} and
entry points of subhaloes \citep{2015ApJ...813....6K, 2018MNRAS.476.1796S}.
\citet{2007MNRAS.375..489H} showed that less massive haloes in clusters
tend to be less spherical and more prolate, and have higher spins
than those in other cosmic environments (void, sheet and filament).
\citet{2011MNRAS.416.1377V} argued that the evolution of halo shape
correlates well with the distribution of the infalling material and
environments.  They suggested that when haloes accrete mass through
narrow filaments at early epochs, they tend to be prolate.  On the
other hand, when the accretion changes more isotropic at later epochs,
the halo becomes oblate or triaxial.

These attempts imply that how haloes accrete mass through filaments is
related to the shape and orientation of them.  This also means how
haloes accrete subhaloes through filaments is important because
subhaloes contribute about $50-70\%$ of the final host halo mass
\citep{2009ApJ...702.1005S}.
These values seem to be consistent with the
picture that about $40\%$ of halo mass comes from smooth accretion of
dark matter particles, which were not bounded by any smaller haloes
\citep{2010ApJ...719..229G}.
The orbit of infalling subhaloes has a highly anisotropic distribution
\citep[e.g.,][]{1997MNRAS.290..411T,2004ApJ...603....7K, 2016ApJ...829...58G},
which would be related to
that the spatial distribution of satellite galaxies around the Milky
Way and the Andromeda galaxy are preferentially aligned in flattened planes
(so-called "plane of satellites")
\citep[e.g.][]{1976MNRAS.174..695L,2005A&A...431..517K,
  2013Natur.493...62I}.  Cosmological simulations have been suggesting
that these anisotropic distributions of satellites may be influenced
by infalling subhaloes along filaments \citep[e.g.,][]{2012MNRAS.421L.137L, 2013MNRAS.428.2489L, 2015ApJ...809...49B}.

\citet{2016ApJ...829...58G} showed that about 20\% number of subhaloes
that were accreted by their host haloes by $z=1$ is coming from
filaments, and this number fraction corresponds to 40\% of the total
subhalo mass, although there is a large halo-to-halo scatter.  The
strength of filamentary accretion of subhaloes would have some impacts
on the shape of haloes as suggested by \citet{2011MNRAS.416.1377V}.
However, \citet{2011MNRAS.416.1377V} used only five haloes, which is
not enough to fully capture the correlation of filamentary subhalo
accretion with the shape and orientation of haloes.

To address these questions, we explore statistics of the correlation
of filamentary subhalo accretion with the shapes (axis ratio) and
orientations of abundant galaxy- and group-sized host haloes with
mass range $5\times10^{11-13}M_{\odot}$, using two high-resolution and
large cosmological $N$-body simulations
\citep{2015PASJ...67...61I,2020MNRAS.492.3662I}.  We identify prime
directions of filamentary subhalo accretion by a similar method
proposed by \citet{2018MNRAS.476.1796S} and calculate the total numbers
and masses of subhaloes accreted along the directions.  The fractions of
them to all subhaloes of each host halo represent the strength of the
anisotropic assembly of the host halo.  Then, we investigate the
impact of anisotropic accretion of subhaloes
on shapes and orientations of host haloes.
This paper provides significant insight for
understanding the galaxy formation histories in the large-scale
universe.

This paper is organized as follows. In Section~\ref{Simulations}, we
describe the details of our two cosmological $N$-body simulations and
sample selection in our work.  In Section~\ref{Methods}, we explain
how we calculate the accretion properties of subhaloes, and shapes and
orientations of their host haloes.  In Section~\ref{Results}, we
present our statistical results of the impact of filamentary accretion of subhaloes
on shapes and orientations of host haloes.
Finally, we discuss and summarize our results in Section~\ref{Disccusion and Sumarry}.

\section{Cosmological $N$-body Simulations}
\label{Simulations}

\begin{table*}
\caption{
Numerical parameters of the two simulations.
$N$ is the number of simulated particles, $L$ is the comoving box size,
$\varepsilon$ is the gravitational softening length, and $m_{\rm p}$ is the particle mass.
}
\centering
\begin{tabular}{cccccc}
\hline
Name & $N$ & $L\ [{\rm Mpc}]$ & $\varepsilon\ [{\rm kpc}]$  & $m_{\rm p}\ [ {\rm M_{\sun}}]$ & reference
 \\
 \hline
 \hline
 $\nu^2{\rm GC}$-H2 & $2048^3$ & 103.0 & 1.57 & $5.06\times10^6$ &\citet{2015PASJ...67...61I}\\
 Phi-1 & $2048^3$ & 47.1 & 0.71 & $4.82\times10^5$  & \citet{2020MNRAS.492.3662I}\\
 \hline
\end{tabular}
\label{tab:tab1}
\end{table*}

We use two large cosmological $N$-body simulations,
the $\nu^2$GC-H2 \citep{2015PASJ...67...61I}
and the Phi-1 \citep{2020MNRAS.492.3662I} summarized in Table \ref{tab:tab1}, and the cosmological
parameters of them are $\Omega_0=0.31, \Omega_{\rm b}=0.048,
\lambda_0=0.69, h=0.68, n_{\rm s}=0.96$, and $\sigma_8=0.83$, which
are consistent with the observation of the cosmic microwave background
obtained by the Planck satellite
\citep{2014A&A...571A..16P,2018arXiv180706209P}.  We identified haloes
and subhaloes by ROCKSTAR \citep{2013ApJ...762..109B} and constructed
their merger trees by CONSISTENT TREES code
\citep{2013ApJ...763...18B}.  The halo/subhalo catalogues and merger
trees of those simulations are publicly available at
\url{http://hpc.imit.chiba-u.jp/~ishiymtm/db.html}.

We analyze galaxy-sized ($M_{\rm host}=5\times10^{11-12}
M_{\sun}$) and group-sized host haloes ($M_{\rm
  host}=5\times10^{12-13} M_{\sun}$) at redshift $z=0$,
where \Mhost\ is the halo virial mass.
We exclude haloes that experience major mergers at $z<1$ from analysis because
recent major mergers are much more likely to affect
structural properties of haloes than continuous accretion of
subhaloes.
We define mergers with the mass ratio greater than 0.3 as
the major merger.
The numbers of host haloes analyzed
in the Phi-1 and the $\nu^2$GC-H2 simulations
are 258 and 2405 for $M_{\rm host}=5\times10^{11-12} \Msun$, respectively,
26 and \textcolor{black}{264} for $M_{\rm host}=5\times10^{12-13} \Msun$, respectively.
We select subhaloes with $M_{\rm \textcolor{black}{ratio}} > 5\times10^{-5}$, where $M_{\rm \textcolor{black} {ratio}}$ is the mass ratio
between their progenitor haloes at the accretion redshift $z_{\rm
  acc}$ and host haloes at $z=0$.  The accretion redshift $z_{\rm
  acc}$ is the time when progenitor haloes first pass through the
virial radius of the most massive progenitors of the host haloes
(so-called `main branch').

\section{Methods}
\label{Methods}

\subsection{The shape and orientation of host haloes}

To quantify the shape and orientation of haloes, we use axis ratios,
$c/a$ and $b/a$ ($a\geqslant b\geqslant c$), and vectors of major
axis $\emajor$, all of which Rockstar computed by the method in
\citet{2006MNRAS.367.1781A}.
The square-roots of the eigenvalues and eigenvectors of the inertia
tensor of haloes correspond to the axis lengths, $a$, $b$ and
$c$, and their vectors, respectively.  We also use angular momentum
vectors $\J$ provided by Rockstar to quantify the orientation of haloes.

\subsection{Filamentary accretion of subhaloes}
To investigate the impact of filamentary accretion of subhaloes on the
shape and orientation of host haloes at $z=0$, we detect
directions from where subhaloes preferentially are accreted.
First, we identify entry points of subhaloes with $M_{\rm ratio} >5\times10^{-4}$ on host haloes, \textcolor{black}{in the order of subhalo mass at $z_{\rm acc}$ based on the previous studies that showed more massive subhaloes are more likely to be accreted into a host halo along a filament \citep[e.g.][]{2014MNRAS.443.1274L,2018MNRAS.481..414G}}.
\textcolor{black}{Here, the entry point of a subhalo is the position of its progenitor halo at $z_{\rm acc}$}.
We regard the direction of entry
point of the most massive subhalo from the centre of host halo as a filament
and assign the remaining subhaloes within the opening angle of
$30^\circ$ from the filament.  Then we apply this procedure to the
next most massive subhalo that is not assigned to any filaments, and
repeat until all the subhaloes are assigned to any filaments.
Finally, we regard the direction of filament $\efilament$
that contains the largest number of assigned subhaloes as the main
filament and use it to quantify the strength of filamentary accretion
of subhaloes.

This method is based on \citet{2018MNRAS.476.1796S}, which quantified
filamentary accretion of subhaloes for their cosmological
hydrodynamical simulations.  The difference is that they applied the
same iterative procedure for the top 11 to 80 massive satellites in
the stellar mass.  Although we use subhaloes with $M_{\rm
  ratio}>5\times10^{-4}$ to detect the main filament, we compared the
results with $M_{\rm ratio}>5\times10^{-5}$ and $M_{\rm
  ratio}>5\times10^{-3}$.  The choice of $M_{\rm ratio}$ does not
strongly affect the directions of main filaments, reinforcing the
effectiveness of our method to detect filamentary accretion of
subhaloes.

\textcolor{black}{
  One may image that two filaments with the first and second largest number of assigned subhaloes should be 180 degrees from each other \textcolor{black}{because accretions preferentially happen along the filaments for haloes in them \citep{2014MNRAS.443.1274L, 2018MNRAS.481..414G}}.
  In our definition of filamentary directions, host haloes with $M_{\rm ratio}=5\times10^{11-12}$ and $5\times10^{12-13}\Msun$, whose cosine of the angle between the two strongest filamentary directions is from -1 to -0.5, are approximately 40\% and 35\%, respectively.
  In case that the directions of two filaments are randomly distributed, only 25\% of their cosine exists in that region, and thus the two strongest filaments tend to be slightly opposite.
  But this trend is weak since cosmic environments haloes reside are not necessarily filaments.
}

We quantify the strength of filamentary accretion of subhaloes for
three groups of subhaloes with different sub-to-host halo mass ratio
ranges of $M_{\rm ratio}=(0.5-5)\times10^{-4},~(0.5-5)\times10^{-3}$, and
$(0.5-5)\times10^{-2}$.  Then we calculate the number and mass fractions
of subhaloes in the main filament relative to all subhaloes in each
group, $F_{\rm number}$ and $F_{\rm mass}$, respectively.  In the case
of high-$F_{\rm number}$ or $F_{\rm mass}$, subhaloes are expected to
be preferentially accreted into their host haloes from a specific
direction, and its accretion is highly anisotropic.  On the other
hand, for haloes with low-$F_{\rm number}$ or $F_{\rm mass}$,
accretion of their subhaloes is expected to be more isotropic than haloes with higher-$F_{\rm number}$ or $F_{\rm mass}$.

\section{Results}
\label{Results}
\subsection{Distributions of number fraction $F_{\rm number}$ and mass fraction $F_{\rm mass}$}
\label{Result-0}

\begin{figure*}
\centering
\includegraphics[width=140mm]{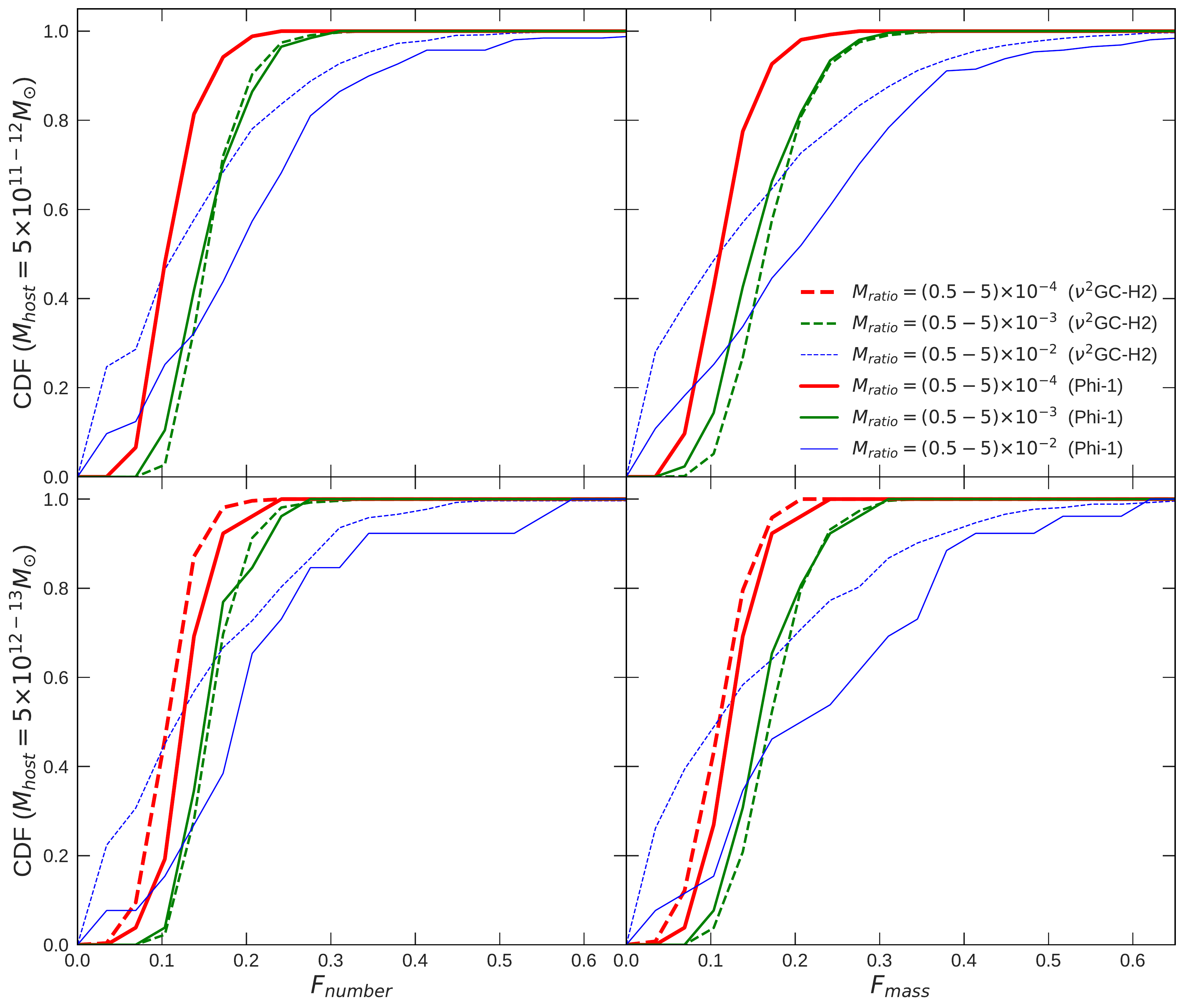}
\caption{
  Cumulative distributions of the number fraction $F_{\rm number}$ (left
  panels) and the mass fraction $F_{\rm mass}$ (right panels) of
  filamentary accretion of subhaloes in host
  haloes with mass ranges of $M_{\rm host}=5\times10^{11-12}M_{\odot}$
  (upper panels) and $M_{\rm host}=5\times10^{12-13}M_{\odot}$ (bottom
  panels). In each panel, dashed and solid curves show the different
  simulations $\nu^2$GC-H2 and Phi-1, respectively.
  Red, green and blue curves with different thickness
  show results with the sub-to-host mass ratio of $M_{\rm
    ratio}=(0.5-5)\times10^{-4},~(0.5-5)\times10^{-3}$ and
  $(0.5-5)\times10^{-2}$, respectively.
  \textcolor{black}{In case that entry points of subhaloes are isotropically distributed, mean fractions with poisson errors for $M_{\rm ratio} = (0.5-5)\times10^{-4}, (0.5-5)\times10^{-3}$ and $(0.5-5)\times10^{-2}$
  are $\sim6.9\pm1.2\%, \sim8.1\pm3.1\%$ and $\sim14.6\pm7.2\%$, respectively.}
  }
\label{fig:1}
\end{figure*}

Fig.~\ref{fig:1} shows the \textcolor{black}{cumulative} distributions of the number fraction $F_{\rm number}$
and the mass fraction $F_{\rm mass}$ of filamentary subhalo accretion
in host haloes with $M_{\rm host}=5\times10^{11-12}$ and $5\times10^{12-13} M_{\sun}$
for three different sub-to-host halo mass ratios,
$M_{\rm ratio}=(0.5-5)\times10^{-4},~(0.5-5)\times10^{-3}$ and $(0.5-5)\times10^{-2}$.
The results of two simulations agree well with each other
regardless of the host halo mass and the sub-to-host mass ratio, \textcolor{black}{except for massive subhaloes with $\Mratio=(0.5-5)\times10^{-2}M_{\sun}$.}

The distributions differ between the sub-to-host mass ratios.
\textcolor{black}{For the sub-to-host mass ratios with $M_{\rm ratio}=(0.5-5)\times10^{-3}$ and $(0.5-5)\times10^{-4}$, 5\% to 95\% of the host haloes have the specific ranges of $0.07\lesssim\Fnumber\, \Fmass\lesssim0.18$ and $0.11\lesssim\Fnumber\, \Fmass\lesssim0.25$, respectively.}
Therefore, host haloes typically accrete \textcolor{black}{$\sim7\%-25\%$} of the total number or mass of
subhaloes with these mass ratios through a specific direction.
\textcolor{black}{These ratios are typically higher than mean fractions with poisson errors for $M_{\rm ratio}=(0.5-5)\times10^{-4}$ and $(0.5-5)\times10^{-3}$, $\sim6.9\pm1.2\%$ and $\sim8.1\pm3.1\%$, respectively, in case that entry points of subhaloes are isotropically distributed.}
\textcolor{black}{The overall distributions of the number fraction and the mass fraction shift higher values with increasing mass ratio, consistent with previous studies that suggested more massive subhaloes come from filament directions \citep{2014MNRAS.443.1274L}.}

Integrating the mass fraction with respect to $M_{\rm ratio}$, we find
that 25 \% on average of the total mass of subhaloes is accreted
through filaments, which is comparable to the result founded by
\citet{2015ApJ...813....6K}.  Considering three filaments that consist
of the three largest number of assigned subhaloes in a host halo, the
average of the mass fraction increases by $\sim40\%$.  This value is
consistent with \citet{2016ApJ...829...58G}, which calculated
the mass fractions of haloes that have an average of three filaments.
These consistencies reinforce the effectiveness of our method to detect
filamentary subhalo accretions, although several different methods
have been proposed to date
\citep[e.g.][and references therein]{2007A&A...474..315A, 2009MNRAS.396.1815F, 2010MNRAS.407.1449G,2011MNRAS.414..384S,2012MNRAS.425.2049H,2014MNRAS.438.3465T, 2014MNRAS.438..177A, 2018MNRAS.473.1195L}.

The fractions of massive subhaloes with $M_{\rm ratio}=(0.5-5)\times10^{-2}$ are widely distributed in the range of
$0.0 \lesssim F_{\rm number}, F_{\rm mass}\lesssim 0.6$.
One of the reason is that the average number of such massive subhaloes is only eleven per host halo,
and thus the halo-to-halo scatter is large.
For such massive subhaloes,
there are some differences between the distributions of $\Fnumber$ and $\Fmass$,
but for less massive ones, as expected, there is no significant difference.
Therefore, although we hereafter show the distributions of halo properties as a function of only $\Fnumber$,
we also confirm similar results as a function of $\Fmass$.
\textcolor{black}{The curves show jumps in the Phi-1 simulation}
for $M_{\rm host}=5\times10^{12-13} \Msun$ because the number of host haloes
is only 26, indicating that the statistics are not enough.
Hereafter, we exclude the Phi-1
simulation for $M_{\rm host}=5\times10^{12-13} \Msun$ from the analysis.

\subsection{Correlation between the host halo shape and filamentary subhalo accretion}
\label{Result-1}

\begin{figure*}
\includegraphics[width=150mm]{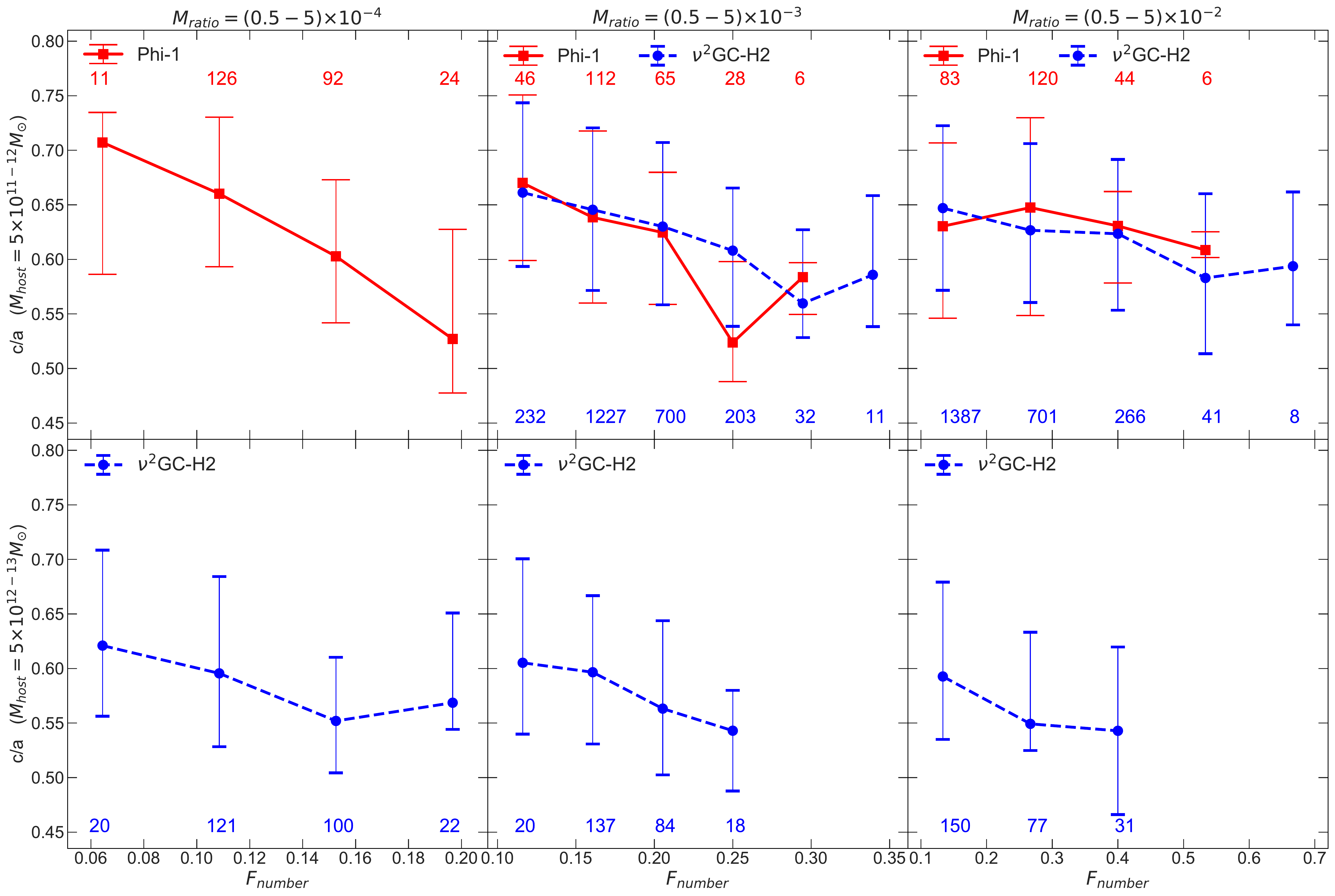}
\caption{
  Axis ratios $c/a$ of host haloes at $z=0$ as a function of the number
  fraction of filamentary accretion $F_{\rm number}$
  for subhaloes with sub-to-host halo mass
  ratios of $M_{\rm ratio}=(0.5-5)\times10^{-4}~{\rm
    (left\ panels)},\ (0.5-5)\times10^{-3}\ {\rm (middle\ panels)}$ and
  $(0.5-5)\times10^{-2}\ {\rm (right\ panels)}$.
  Upper and lower panels are for the host halo mass ranges of $M_{\rm host}=5\times10^{11-12}M_{\sun}$
  and $M_{\rm host}=5\times10^{12-13}M_{\sun}$, respectively.
 Red solid and blue dashed curves are results from the Phi-1 and the $\nu^2$GC-H2
  simulations, respectively. Squares and circles are the median values, and whiskers
  are the first and third quantiles in each $F_{\rm number}$
  bin. The number of haloes in each bin for the Phi-1 and the $\nu^2$GC-H2 simulations are
  specified at the upper and bottom of whiskers, respectively.
  We only
  plot bins in which the number of haloes is greater than
  five. The results with $M_{\rm host}=5\times10^{11-12}M_{\sun}$ and
  $M_{\rm ratio}=(0.5-5)\times10^{-4}$ of the
  $\nu^2$GC-H2 simulation are not displayed because of its mass
  resolution limit.
}
\label{fig:2}
\end{figure*}

Fig.~\ref{fig:2} shows the axis ratios $c/a$ of host haloes
with $M_{\rm host}=5\times10^{11-12}$ and $5\times10^{12-13} M_{\sun}$
at $z=0$ as a function of the number fraction of filamentary accretion $F_{\rm number}$
for three different sub-to-host mass ratios,
$M_{\rm ratio}=(0.5-5)\times10^{-4},~(0.5-5)\times10^{-3}$ and
$(0.5-5)\times10^{-2}$.
We find that axis ratios depend on the number fraction
regardless of the host halo mass and the
sub-to-host mass ratio.
The shape of haloes tends to be more elongated with increasing anisotropy of
subhalo accretion, although the correlation is rather weak.
For haloes with $M_{\rm host}=5\times10^{11-12}$ and $5\times10^{12-13} M_{\sun}$
in the $\nu^2$GC-H2 simulation,
the median values of $c/a$ are $\sim0.64$ and $\sim0.58$, respectively,
which exist between the first and third quantiles for most $M_{\rm ratio}$ bins.

This weak correlation between $F_{\rm number}$ and $c/a$ becomes
slightly stronger with decreasing mass ratio $M_{\rm ratio}$.  We do
not show the result of the Phi-1 simulation for $M_{\rm
  host}=5\times10^{12-13}$ because the statistics are not enough.
However, the existence of the correlation is supported by the $\nu^2$GC-H2
simulation.  The median axis ratios $c/a$ tend to be lower in
more massive haloes than in less massive haloes, 
consistent with previous studies \citep[e.g.][]{2006MNRAS.367.1781A}.

\begin{figure}
\includegraphics[width=90mm]{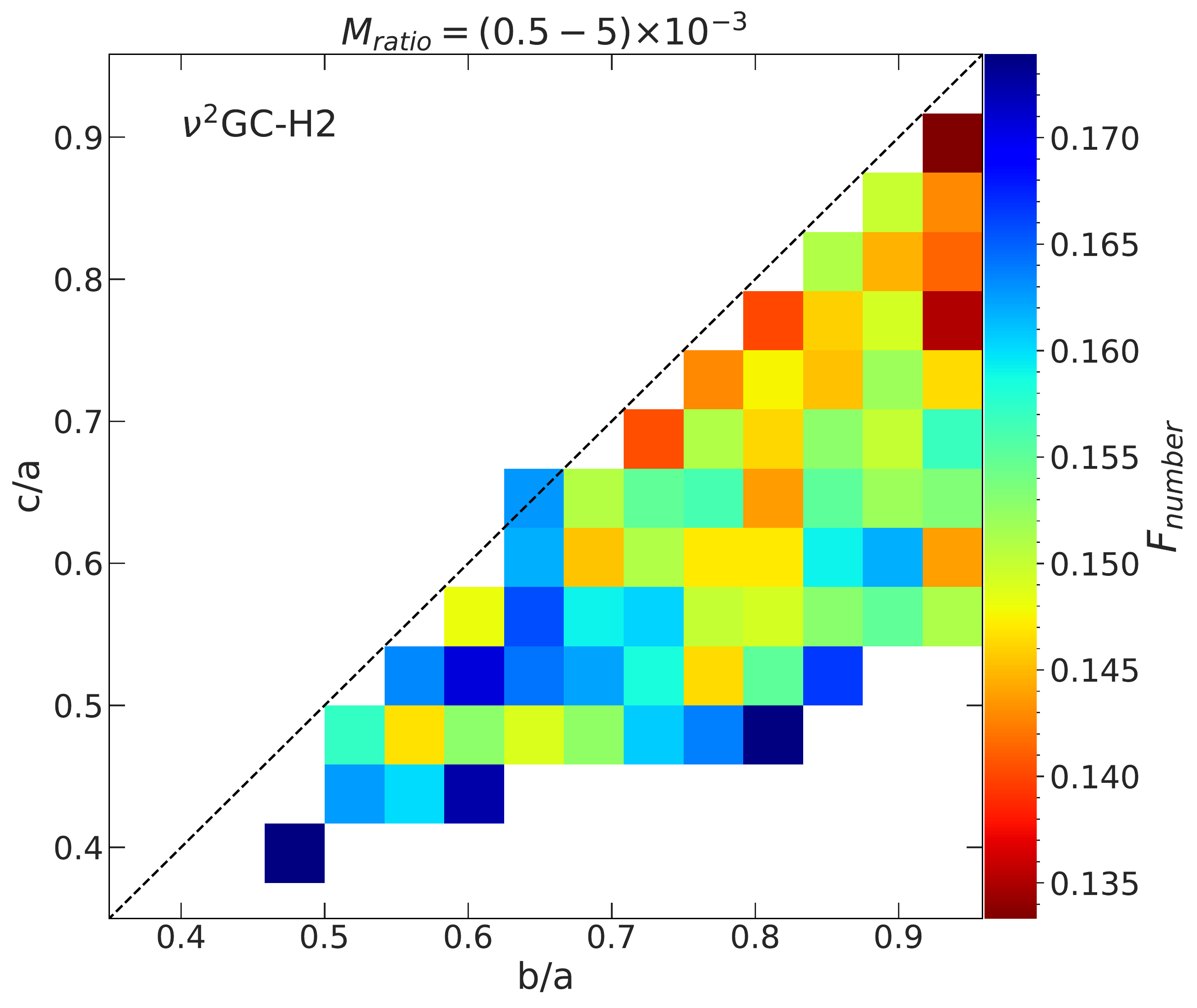}
\caption{
  Distribution of axis ratios $b/a$ and $c/a$ of host haloes with
  $M_{\rm host}=5\times10^{11-12}M_{\sun}$ in the $\nu^2$GC-H2 simulation
  as a function of number fraction of filamentary accretion $F_{\rm number}$
  for $M_{\rm ratio}=(0.5-5)\times10^{-3}$.  Colour represents the median value of
  $F_{\rm number}$ for haloes in each region,
  in which the number of haloes is greater than five.
}
\label{fig:3}
\end{figure}

To see the relationship between the shapes of host haloes
and anisotropic subhalo accretion more precisely, we plot the
distributions, $b/a$ versus $c/a$, as a function of $F_{\rm number}$ in Fig.~\ref{fig:3}.
The $c/a$ and $b/a$ vary smoothly as the number fraction of filamentary accretion
$F_{\rm number}$, and haloes with low-$F_{\rm number}<0.15$
tend to be more spherical or oblate ($c/a>0.7$ and $b/a>0.8$).
On the other hand, with increasing $F_{\rm number}$,
haloes tend to become more prolate or triaxial with $c/a\sim0.4-0.6$
and $b/a\sim0.5-0.9$ at $z=0$.
These trends are mentioned in \citet{2011MNRAS.416.1377V},
which analyzed only five Milky Way-sized haloes.
Here, we first quantify these trends statistically
and show the correlation between the halo shape and
filamentary subhalo accretion, with a number of haloes.

\subsection{Accretion alignment with the orientation of host halo major axis}
\label{Result-2}

\begin{figure*}
\includegraphics[width=150mm]{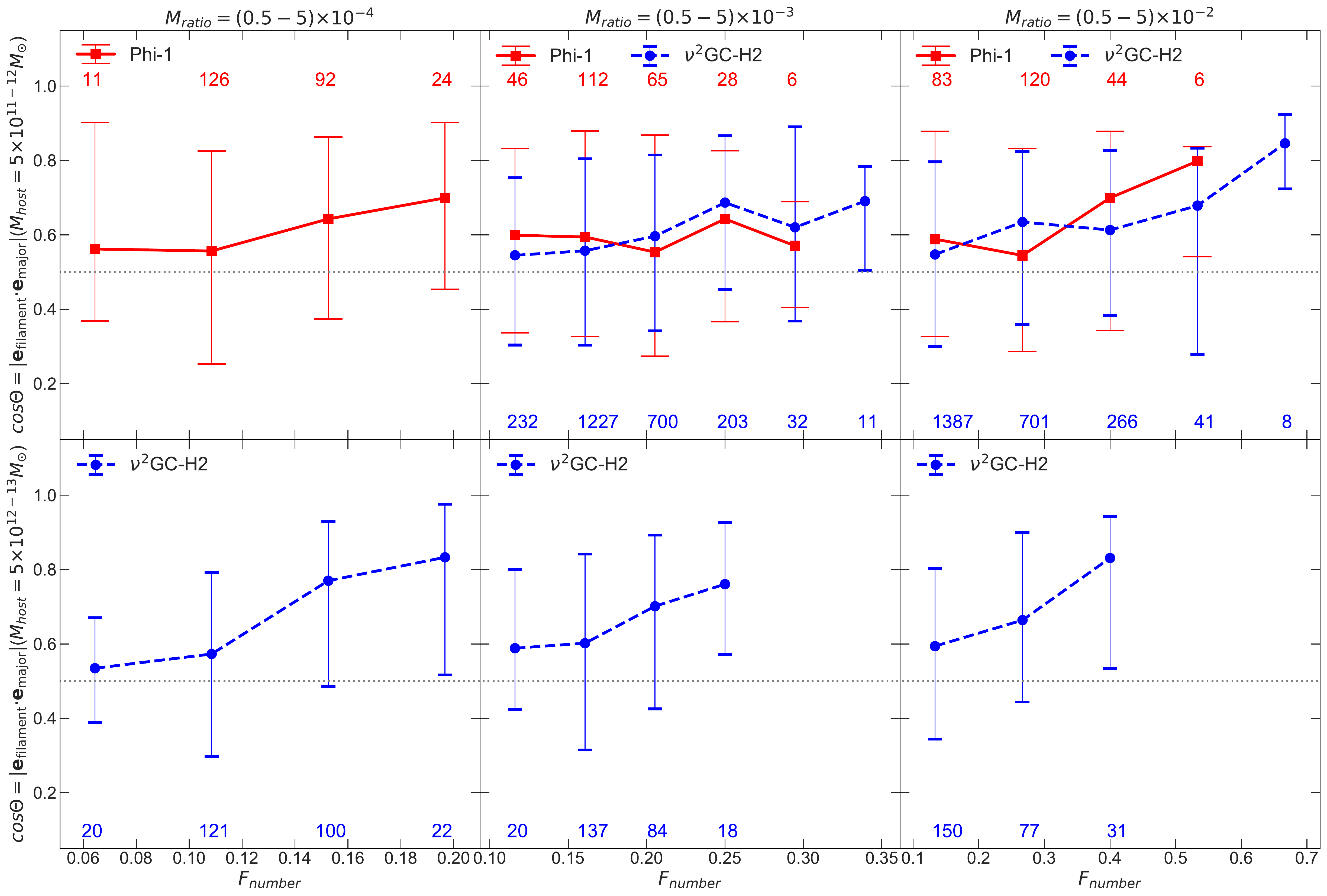}
\caption{
Cosine of the angle between the filamentary accretion directions
$\efilament$ and the orientations of host halo major axes $\emajor$
at $z=0$ as a function of the number fraction $F_{\rm number}$, in the
same format as Fig.~\ref{fig:2}.  Dotted horizontal lines at
cos$\theta=0.5$ correspond to the random distribution of angle between
$\efilament$ and $\emajor$.
}
\label{fig:4}
\end{figure*}

Fig.~\ref{fig:4} shows the cosine between the filamentary accretion
directions $\efilament$ and the orientations of host halo major
axis $\emajor$ at $z=0$ as a function of the number fraction of
filamentary accretion $F_{\rm number}$.
Two simulations results agree well with each other.
The median values of cosine are higher than 0.5 in
any cases, $\sim 0.57-0.60$, signifying that the major axis tends
to be preferentially aligned with the filamentary accretion
direction.
These results are consistent with previous studies that
showed the orientations of major axes tend to be a little alighted with the
directions of filaments \citep{2007MNRAS.381...41H,
  2009ApJ...706..747Z, 2013MNRAS.428.2489L} and entry points of
subhaloes \citep{2015ApJ...813....6K,2018MNRAS.476.1796S}.  We also
find that the cosine increases with increasing the number fraction
$F_{\rm number}$, indicating that the major axis with higher
anisotropic subhalo accretion tends to be more aligned with the
filamentary accretion direction.
We can see that these trends do not clearly depend on
the sub-to-host halo mass ratio.

The correlation is slightly stronger in more massive host haloes than
in less massive ones.  Besides, major axes of more massive host
haloes tend to be slightly more aligned with filamentary directions on
average than those of less massive ones,
consistent with \citet{2018MNRAS.481..414G}.
Subhaloes are preferentially accreted along major axes of
host haloes, and this trend is stronger with increasing host halo masses
\citep{2015ApJ...813....6K}. Therefore, halo mass dependence seen in
\fref{fig:4} reflects these trends found by previous studies.

\subsection{Accretion alignment with the orientation of host halo angular momentum vector}
\label{Result-3}

\begin{figure*}
\includegraphics[width=150mm]{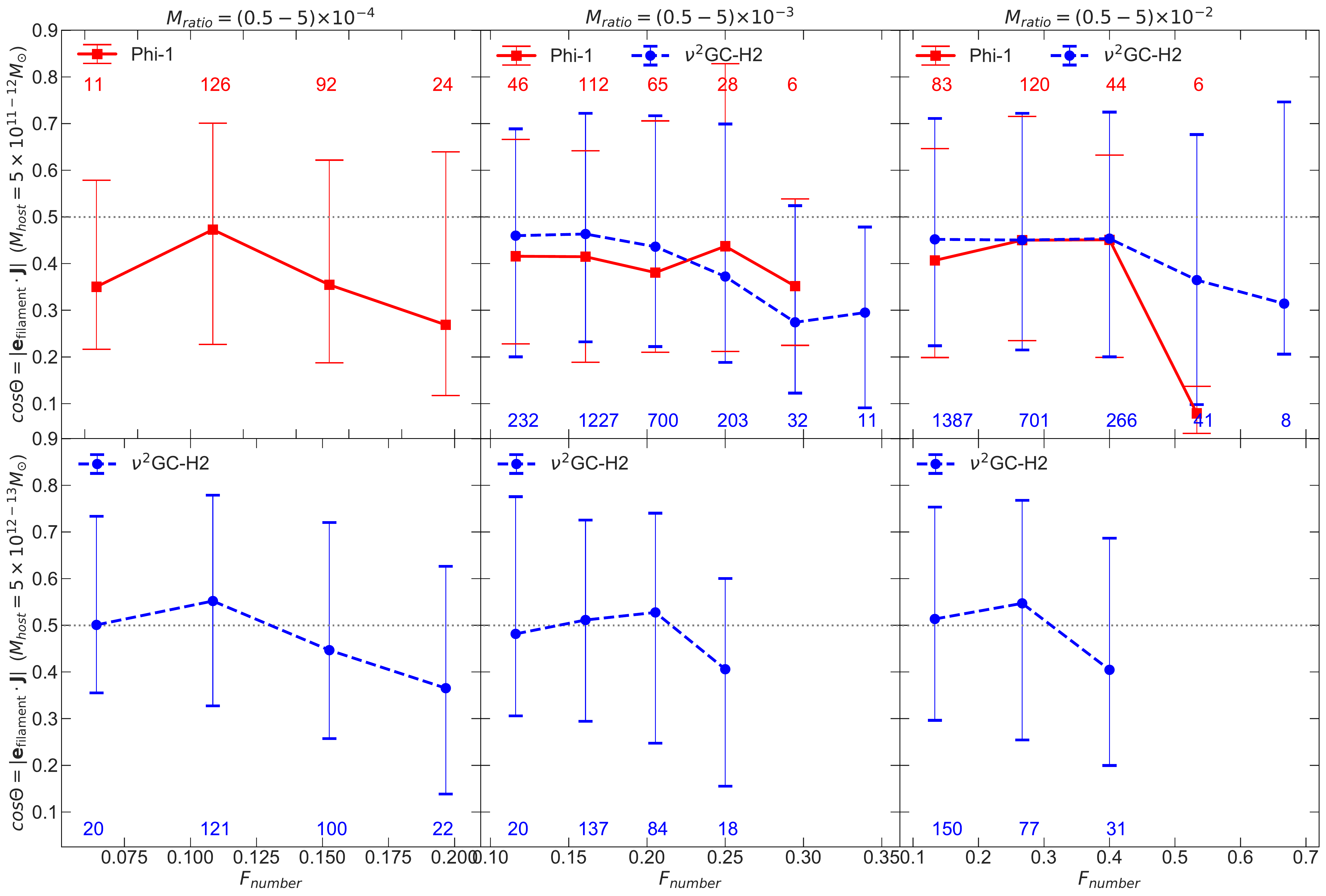}
\caption{
Cosine of the angle between the filamentary accretion directions
$\efilament$ and the vectors of host halo angular momentum $\bm{J}$
at $z=0$ as a function of the number fraction $F_{\rm number}$, in
the same format as Fig.~\ref{fig:2}.  Dotted horizontal lines at
cos$\theta=0.5$ correspond to the random distribution of angle between
$\efilament$ and $\bm{J}$.
}
\label{fig:5}
\end{figure*}

Fig.~\ref{fig:5} shows the cosine between the filamentary accretion
directions $\efilament$ and the host halo angular
momentum vectors $\J$ at $z=0$ as a function of the number fraction of
filamentary accretion $F_{\rm number}$.
Two simulations results agree well with each other, except for the massive end for
$M_{\rm host}=5\times10^{11-12}M_{\sun}$ and $M_{\rm
  ratio}=(0.5-5)\times10^{-2}$, probably due to the lack of statistics
of the Phi-1 simulation.
For $M_{\rm host}=5\times10^{11-12}M_{\sun}$, the median values of
cosine are lower than 0.5 in any cases, $\sim 0.45-0.47$.  This
indicates that the angular momentum vectors $\J$ tend to be
preferentially perpendicular to the directions of filaments,
consistent with previous studies \citep[e.g.][]{2012MNRAS.421L.137L}
and the assumption that the angular momentum
of haloes results from the transfer of the orbital angular
momentum of subhaloes accreted along filamentary accretion directions.

\fref{fig:5} shows that the cosine decreases with increasing the
number fraction $F_{\rm number}$, indicating that the angular momentum
vector with higher anisotropic subhalo accretion tends to be slightly
more misaligned with the filamentary accretion direction.
We can see that these trends do not clearly depend on
the sub-to-host halo mass ratio.
These correlations become weaker in more massive haloes than in less massive
haloes. For $M_{\rm host}=5\times10^{12-13}M_{\sun}$, the median
values of cosine are around 0.5, which corresponds to the random
distribution.
\citet{2005ApJ...627..647B} argued that the angular momentum
of more massive host haloes are mainly built up by recent accretion of a
few massive subhaloes along filaments, while that of less massive ones are
built up by smooth accretion of subhaloes.
In this work, $\Fnumber$ represents the strength of smooth accretion of subhaloes
rather than the contribution of a few massive subhaloes.
These factors  would explain why
the cosine for more massive haloes does not show a clear correlation
with $\Fnumber$ and distributes somewhat randomly.

\subsection{\textcolor{black}{Alignment between the shape and the angular momentum vector of host halo}}
\label{Result-4}

\begin{figure*}
\includegraphics[width=150mm]{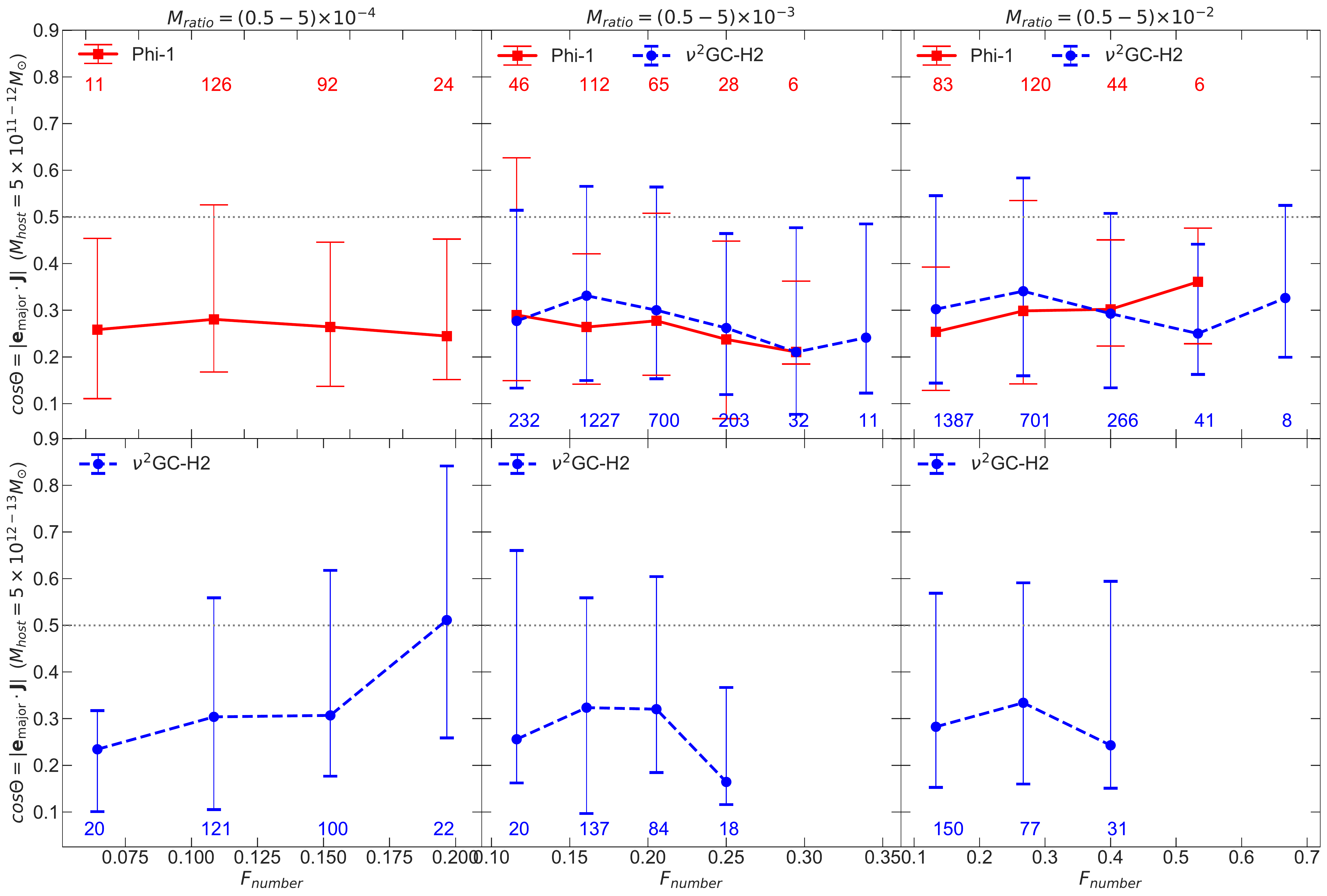}
\caption{
\textcolor{black}{Cosine of the angle between the orientation of host halo major axes $\emajor$ and the vectors of host halo angular momentum $\bm{J}$
at $z=0$ as a function of the number fraction $F_{\rm number}$, in
the same format as Fig.~\ref{fig:2}.  Dotted horizontal lines at cos$\theta=0.5$ correspond to the random distribution of angle between $\efilament$ and $\bm{J}$.}
}
\label{fig:6}
\end{figure*}

\textcolor{black}{
  Fig.~\ref{fig:6} shows the cosine between the orientation of host halo major axes $\emajor$ and the vectors of host halo angular momentum $\J$ at $z=0$ as a function of the number fraction of filamentary accretion $F_{\rm number}$.
  Although there is no significant correlation between cosine and $\Fnumber$, almost all median values of cosine are lower than 0.5 in any cases, $\sim$0.2-0.3, signifying that the major axes of host haloes tend to be perpendicular to their angular momentum.
  This trend is consistent with previous studies \citep[e.g.][]{2006MNRAS.367.1781A,2007MNRAS.376..215B} that
  showed the angular momentum of haloes tends to be parallel to their minor axes and to be perpendicular to their major axes.
  We also find that the distributions of the cosine do not depend on the host halo mass and sub-to-host halo mass ratio.
}

\section{Discussion and Summary}
\label{Disccusion and Sumarry}

Previous studies suggested that the shape and orientation of host haloes correlate with some properties of
filaments, with significant dependence on the halo mass.
For example, major
axes of haloes are more aligned with filaments as the halo
mass increases \citep{2009ApJ...706..747Z, 2015ApJ...813....6K,
  2016MNRAS.460.3772S}, and angular momentum vectors of less massive
haloes ($\lesssim10^{12} M_{\sun}$) tend to be slightly parallel to the filaments while those of
more massive haloes ($\gtrsim10^{12} M_{\sun}$) tend to be perpendicular
\citep{2007ApJ...655L...5A, 2007MNRAS.381...41H,
  2013MNRAS.428.2489L,2018MNRAS.481..414G}.
In this paper, we have extended these studies in terms of filamentary
accretion of subhaloes and have investigated the impact of them
on the shape and orientation of haloes.

Using large cosmological $N$-body simulations, we have analyzed a
large number of galaxy-sized haloes ($\Mhost=5\times10^{11-12}\Msun$)
and group-sized haloes ($\Mhost=5\times10^{12-13}\Msun$), which enable
us to compare the statistics in structural properties for host haloes
with different accretion histories.  Detecting entry points of
subhaloes by tracing their progenitor orbits, we have identified the
main filaments and their directions from the centre of host haloes.
To quantify the strength of filamentary subhalo accretion with
different sub-to-host mass ratios,
$M_{\rm ratio}=(0.5-5)\times10^{-4},~(0.5-5)\times10^{-3}$ and
$(0.5-5)\times10^{-2}$, we have calculated the number fraction $\Fnumber$ and mass
fraction $\Fmass$ of subhaloes in the main filament relative to all subhaloes
for each mass ratio.

We have confirmed the host halo mass dependence of the alignment
between orientations of haloes and filaments found by previous
studies.  We have also shown that, for the first time, the shape and
orientation of haloes at $z=0$ weakly correlate with the strength
of filamentary subhalo accretion even if the host halo masses are the
same.  The minor-to-major axis ratio $c/a$ and intermediate-to-major
axis ratio $b/a$ significantly depend on the strength
of filamentary subhalo accretion regardless of the host halo mass and
the sub-to-host mass ratio.  The shape of haloes tends to be more
elongated with increasing number fraction $\Fnumber$.  Moreover, host
haloes with highly anisotropic accretion become more spherical or oblate
($c/a>0.7$ and $b/a>0.8$), while host haloes with isotropic accretion
become more prolate or triaxial ($c/a\sim0.4-0.7$ and
$b/a\sim0.5-0.9$).

With increasing the number fractions of subhaloes from filamentary
accretion directions, the major axes are preferentially aligned with
the directions, while their angular momentum vectors tend to be
slightly more misaligned with the directions.  On the other hand, with
decreasing the number fractions of subhaloes from filamentary accretion
directions, their alignment angles with the major axes and the angular
momentum vectors tend to be randomly distributed.

These correlations are seen in haloes with $\Mhost=5\times10^{11-12}$
and $5\times10^{12-13}\Msun$, and their strength is
slightly different.  This halo mass dependence is expected to result
from different infall patterns of subhaloes with different host halo
mass.  On the other hand, there is no significant difference in the
distributions of the shape and orientations as a function of number
fraction for the different sub-to-host mass ratio of
$\Mratio=(0.5-5)\times10^{-4},~(0.5-5)\times10^{-3}$ and
$(0.5-5)\times10^{-2}$.

Our studies have been highlighting that the shape and orientation of
haloes correlate with not only large-scale cosmic environments but the
strength of filamentary subhalo accretion.
This implies that, as seen in intrinsic
alignments of galaxies and haloes \citep{2005ApJ...627..647B,2013MNRAS.428.1827T,2017ApJ...848...22X,2019PhRvD.100j3507O},
large-scale cosmic environments would leave some imprints on stellar
streams, galaxies and their satellites.
Their observational signatures are influenced by how haloes accrete subhaloes,
such as the plane of satellites \citep{2012MNRAS.421L.137L, 2013MNRAS.428.2489L},
and statistics of streams \citep{2019MNRAS.487.2718M}.
In future studies, we will perform theoretical studies to connect these signatures and large-scale cosmic environments and
compare with observations provided by new facilities, which help to
understand galaxy formation histories over cosmic time.

\section*{Acknowledgements}
We thank Takanobu Kirihara for valuable comments and discussions.
We also thank an anonymous referee for helpful comments.
Numerical computations were partially carried out on the K computer at the RIKEN Advanced
Institute for Computational Science (Proposal numbers hp150226,
hp160212, hp170231, hp180180), Aterui and Aterui II supercomputer at
Center for Computational Astrophysics, CfCA, of National Astronomical
Observatory of Japan.  This work has been supported by MEXT as
``Priority Issue on Post-K computer'' (Elucidation of the Fundamental
Laws and Evolution of the Universe) and JICFuS. We thank the support
by MEXT/JSPS KAKENHI Grant Number 17H04828, 17H01101 and 18H04337.

\bibliographystyle{mnras}
\bibliography{example}

\bsp	
\label{lastpage}
\end{document}